\documentclass[letterpaper, 10 pt, conference]{ieeeconf}  

\usepackage{booktabs}
\usepackage{nccmath, mathtools}
\usepackage{graphicx}   
\usepackage{xcolor}
\usepackage{subcaption}
\usepackage{wrapfig,lipsum}
\usepackage{amsmath,amssymb,amsfonts}
\usepackage[font=footnotesize,labelfont=bf]{caption}
\usepackage{nomencl}
\makenomenclature
\IEEEoverridecommandlockouts                              

\begin{document}
\title{\LARGE \bf
Efficient Planning of Multi-Robot Collective Transport using Graph Reinforcement Learning with Higher Order Topological Abstraction
}
\author{Steve Paul$^{1}$, Wenyuan Li$^{2}$, Brian Smyth$^{3}$, Yuzhou Chen$^{4}$, Yulia Gel$^{5}$, and Souma Chowdhury$^{6,\dagger}$
\thanks{$^\dagger$ Corresponding Author, soumacho@buffalo.edu}
\thanks{Authors $^{1}, ^{2}, ^{3}, ^{6}$ are with the Department of Mechanical and Aerospace Engineering, University at Buffalo, Buffalo, NY, USA {\tt\small \{stevepau, wli3535, briansmy, soumacho\}@buffalo.edu}}
\thanks{Author $^{4}$ is with the Department of Computer and Information Sciences,
Temple University, Philadelphia, PA, USA
{\tt\small yuzhou.chen@temple.edu}}
\thanks{Author $^{5}$ is with the Department of Mathematical Sciences,
University of Texas at Dallas, Dallas, TX, USA
{\tt\small ygl@utd.edu}}
\thanks{This work was supported by the Office of Naval Research (ONR) award N00014-21-1-2530 and the National Science Foundation (NSF) award CMMI 2048020. Any opinions, findings, conclusions, or recommendations expressed in this paper are those of the authors and do not necessarily reflect the views of the ONR or the NSF.}
\thanks {\copyright\space2023 IEEE. Personal use of this material is permitted. Permission from IEEE must be obtained for all other uses, in any current or future media, including reprinting/republishing this material for advertising or promotional purposes, creating new collective works, for resale or redistribution to servers or lists, or reuse of any copyrighted component of this work in other works.}
}

\maketitle
\thispagestyle{empty}
\pagestyle{empty}

\begin{abstract}
Efficient multi-robot task allocation (MRTA) is fundamental to various time-sensitive applications such as disaster response, warehouse operations, and construction. This paper tackles a particular class of these problems that we call MRTA-collective transport or MRTA-CT -- here tasks present varying workloads and deadlines, and robots are subject to flight range, communication range, and payload constraints. For large instances of these problems involving 100s-1000's of tasks and 10s-100s of robots, traditional non-learning solvers are often time-inefficient, and emerging learning-based policies do not scale well to larger-sized problems without costly retraining. To address this gap, we use a recently proposed encoder-decoder graph neural network involving Capsule networks and multi-head attention mechanism, and innovatively add topological descriptors (TD) as new features to improve transferability to unseen problems of similar and larger size. Persistent homology is used to derive the TD, and proximal policy optimization is used to train our TD-augmented graph neural network. The resulting policy model compares favorably to state-of-the-art non-learning baselines while being much faster. The benefit of using TD is readily evident when scaling to test problems of size larger than those used in training.
\end{abstract}

\section{Introduction \& Motivation}\vspace{-.18cm}

Efficient solutions based on the use of multi-robot teams show increasing promise in a variety of applications ranging from disaster response \cite{ghassemi2021multi} to manufacturing \cite{8785546}, warehouse logistics \cite{xue2019task} and construction \cite{meng2008distributed}. In most large-scale applications involving 100's to 1000's of tasks, using a centralized command center to perform task assignments is unlikely to be robust due to communication limitations, a single point of failure, and the likelihood of information overloading on one command center. This leads to the need for robots to take decisions in a decentralized yet timely (real-time) manner. In addition, key problem complexities include tasks with deadlines, and heterogeneity of tasks in terms of demand or workload (e.g., requiring one vs. multiple robots or trips) -- such features are commonplace in the stated application scenarios. Moreover, we must account for other practical constraints, namely robot range, robot capacity (e.g., payload capacity), and limited communication range of each robot. 

Given this context, in this paper, we focus on a class of multi-robot task allocation (MRTA) problems that we call MRTA-Collective Transport (MRTA-CT). MRTA-CT involves using a team of robots to perform tasks that are spatially distributed, and present time deadlines and different workloads that may require multiple visits by robots to complete the task. We posit that this problem scenario generalizes to a wide range of material transport applications, which are discussed later in this section. In addition, to introduce a sufficient degree of realism, we consider that decisions must be taken in a decentralized asynchronous manner by each robot, with robots having partial observability about the state of peer robots and the state of tasks as they start to get completed (due to communication range limitations). In addition, for ease of implementation, we assume that the robot team has full observability of tasks at the start of the operation, and a single depot (material source and recharging location) is used by the entire team. 

\textbf{Motivating examples:} This paper focuses on MRTA-CT problems that arise in real-world operations such as: \textbf{I)} Disaster relief operations e.g., in a flood response scenario \cite{ghassemi2021multi}, where varying amounts of relief packages from a central depot must be time-efficiently delivered to victims stranded in a spatially distributed manner over the region; \textbf{II)} Manufacturing or construction sites, where a processed entity has to be delivered from a single source to multiple locations over a large site, based on their varying demand. For disaster relief, strict time deadlines are self-evident. In both scenarios, time constraints are crucial and tasks must be completed before specific deadlines. These deadlines are considered hard constraints, meaning that tasks are only considered completed if their entire demand is met before the deadline. The participating robots have a maximum payload capacity and range due to battery or fuel limitations. Any robot can partially fulfill the demand of each task location as long as the deadline has not passed. For example, if a location requires 10 relief kits and a robot can carry 5, multiple robots can deliver the kits, and a robot with remaining capacity can complete another task. If a robot carrying its full payload selects a task that only needs a few more relief kits, it can deliver them and select another task if it has enough range to do so. This approach can be used in various real-world applications, where products or raw materials must be delivered within specific timeframes for the timely execution of spatially distributed jobs that depend on those products.


\vspace{-.2cm} \subsection{Related Works:} \vspace{-.2cm}
While very few methods exist to directly tackle this particular MRTA-CT class of problems, there's a rich body of work on related methods in MRTA that could potentially be transitioned to this class. These methods, namely graph-matching methods~\cite{ghassemi2021multi,ismail2017decentralized}, mixed integer-linear programming (MILP) approaches~\cite{nallusamy2009optimization,toth2014vehicle}, and auction-based methods~\cite{dias2006market,schneider2015auction}) typically aim to solve the \textbf{\textit{combinatorial optimization (CO)}}  problem underlying MRTA planning. However, these methods often do not scale well with the number of robots and/or tasks and do not readily adapt to complex problem characteristics without tedious hand-crafting of the underlying heuristics. 

In recent years, reinforcement learning or RL methods that use Graph Neural Networks (GNN) are being increasingly used to solve such planning problems with a CO formulation ~\cite{Kool2019, barrett2019exploratory,doi:10.2514/6.2022-3911, khalil2017learning, Kaempfer2018LearningTM, 9750805, li2018combinatorial, nowak2017note, 9116987, Tolstaya2020MultiRobotCA, Sykora2020, Dai2017}. This emergence of graph RL is partly attributed to the ability of GNNs to capture both Euclidean and non-Euclidean features along with local and global structural information of the task space. These methods are, however, limited in three key aspects: \textbf{1)} Simplified problems that often exclude common real-world factors such as resource and capacity constraints \cite{Kool2019, Kaempfer2018LearningTM, khalil2017learning, Tolstaya2020MultiRobotCA}). \textbf{2)} Focused on smaller sized problems ($\leq$ 100 tasks and 10 robots)~\cite{Malcolm2005, 9116987}. \textbf{3)} Rarely provide evidence of generalizing to problem scenarios that are larger in size than those used for training. Such capability would be particularly critical since real-world MRTA problems often involve simulating episodes whose costs scale with the number of tasks and robots, making re-training efforts burdensome. For practical scenarios with large numbers of task locations, achieving good feasible solutions is typically the priority \cite{Cappart2021}, especially under constrained communication scenarios \cite{Cao10complextasks} -- which further motivates the need for learning-based policies to drive real-time planning in such applications.


To enable scalable policies that can be executed in real-time, a novel encoder-decoder-based RL approach was introduced by our earlier work \cite{Paul_ICRA, osti_10345362}. We now hypothesize that scalability can be further improved by utilizing task-neighborhood similarity. To this end, we use the Capsule Attention Mechanism or \textbf{CAPAM} policy network introduced in \cite{Paul_ICRA,osti_10345362}, and particularly augment it with \textit{Topological Descriptors} (TD) as novel additional task-space features to compute task-neighborhood similarity. The policy network is trained using a standard policy gradient RL algorithm, Proximal Policy Optimization (PPO) \cite{schulman2017proximal}. 

Topological Data Analysis (TDA) involves the extraction of higher-order shape features from an observed object such as graph-structured data. By shape here we broadly understand object properties that are invariant under continuous transformations such as bending, twisting, and compressing.
It relies on the intuition that the extracted shape characteristics contain some inherent hidden information on the underlying object that enhance learning capabilities. The primary TD used here is Persistence Diagrams (PD) \cite{cohen2005stability}. 
The new proposed policy network is then trained to learn sequential actions for MRTA-CT (meaning task selection by each robot taking decisions) from an output probability distribution with the overall objective to maximize the number of tasks completed. 
For taxonomy purposes, the MRTA-CT problem can be classified as Single-task Robots, Multi-robot Tasks, Time-extended Assignment (ST-MR-TA) class defined in \cite{gerkey2004formal,nunes2017taxonomy}, which is an $\mathcal{NP}$-hard problem. Based on iTax taxonomy as defined in \cite{korsah2013comprehensive}, this problem also falls into the In-schedule Dependencies (ID) category. The optimization formulation of the MRTA-CT as defined in this paper yields a large mixed-integer non-linear programming (MINLP) problem, which further elucidates the substantial problem complexity.

\textbf{Key Contributions:} 
The main contributions of this paper can be summarized as \textbf{1)} Formulating the MRTA-CT problems as a Markov Decision Process or MDP over graphs such that the task allocation policy can be learned using a policy gradient RL approach, where the task information is represented as node embeddings from a GNN-based encoder of the policy network, the robot states embedded as the \textit{context} portion of the policy network, and the above two information is used to select the next task using the Attention-based decoder in a sequential manner. \textbf{2)} Explore the advantage of using higher-order structural information (encoded by TD) as additional features in improving the generalizability and scalability of the GNN-based policies. \textbf{3)} Demonstrate this learning framework's ability to generalize to larger-sized problems without the need to retrain.

The next section briefly overviews the MRTA-CT problem and its MDP formulation. Section \ref{sec:Learning_Framework} describes our proposed new graph learning architecture that operates on this MDP. Section \ref{sec:Experimental_Evaluation} presents the settings and outcomes of numerical experiments performed on MRTA-CT problems of varying size, used for comparative evaluation of the learning and non-learning methods, and MINLP (optimal) solutions. Section \ref{sec:Conclusion} summarizes concluding remarks and potential future extensions of our work. 
\vspace{-.15cm}

\section{MRTA - Collective Transport}\vspace{-.2cm}

\label{sec:Multi_Robot_collective_Transport}
\subsection{Problem description}\label{subsec:problem_description}
\begin{figure}[!ht]
\vspace{-.35cm}
    \centering
    
    \includegraphics[width=0.42\textwidth]{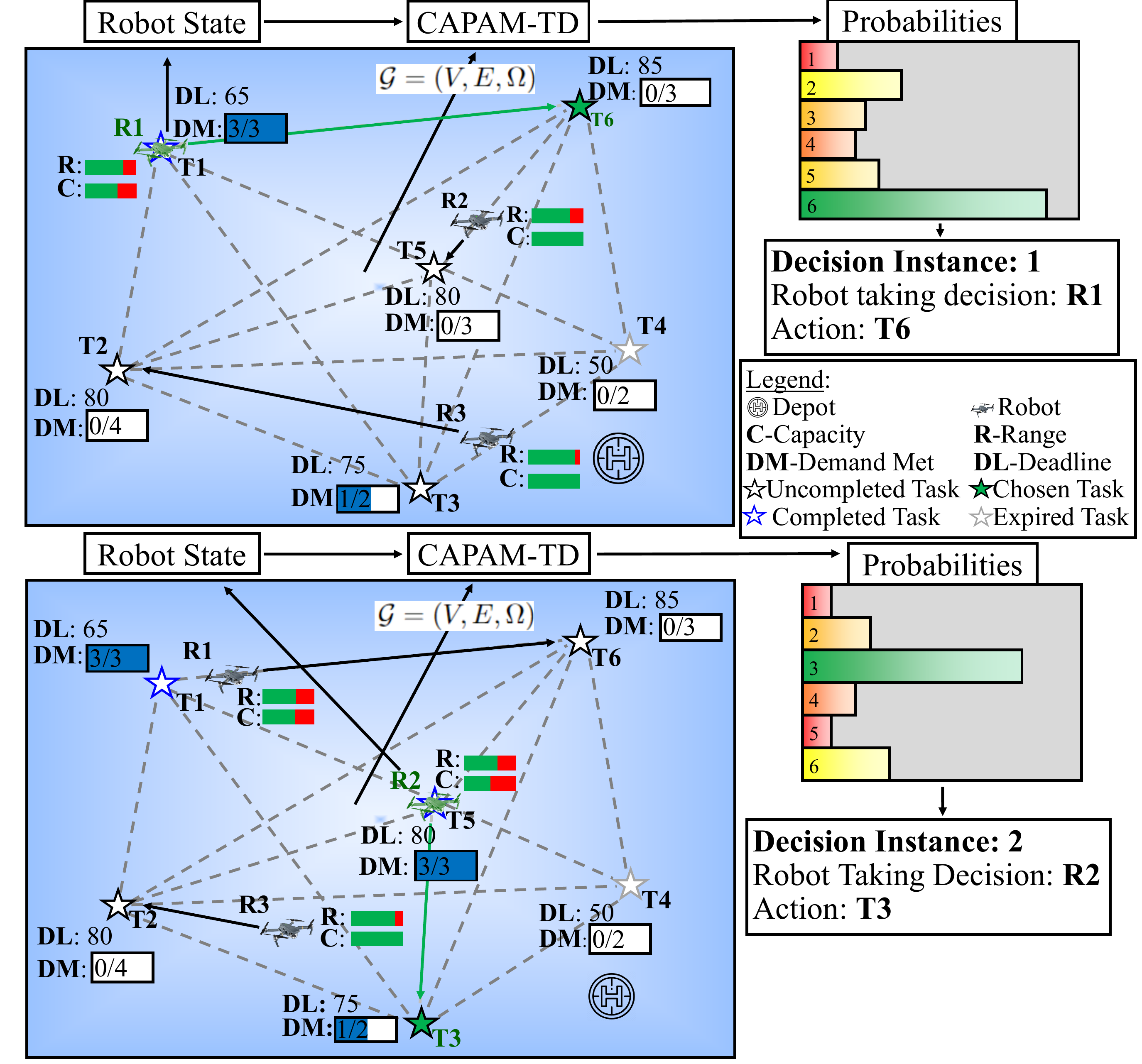}
    \vspace{-0.1cm}
    \footnotesize
    \caption{Two sequential decision-making instants in MRTA-CT problem with 6 tasks and 3 robots. The task with the largest probability is chosen.}
    \label{fig:decision_making}
    \vspace{-.4cm}
\end{figure}

\vspace{-.18cm}
Given a homogeneous set of $M$ robots, $R$=$\{r_{1},r_{2},\ldots,r_{M}\}$ and a set of $N$ tasks $V$, the goal is to allocate tasks to robots for maximizing a given objective function. The objective here is to maximize the number of tasks done. There is a single depot that serves as the start and end points of each robot. Each task $i \in V$ has a unique location represented by its x-y coordinates ($x_{i}, y_{i}$), a workload/demand $w_{i}$ (time-varying) which can be fulfilled partially by a robot, and a time deadline $\tau_i$ by which the task must be completed (demand satisfied) to consider the task $i$ as \texttt{done} (namely $\rho_i$=$1$). Fig~\ref{fig:decision_making} illustrates the MRTA-CT problem. Each robot has a maximum distance range, $\Delta_{\text{max}}$, that it can travel before returning to the depot to recharge; each robot also has a defined maximum capacity $C_{\text{max}}$. Each robot starts from the depot where it gets a full battery and full load and then visits the task location to satisfy (fully or partially) its demand. A robot returns to the depot once it is fully unloaded, it is running out of battery, or there are no more remaining tasks in the environment, whichever comes first. The recharging process is assumed to be instantaneous (e.g., via battery swap).  \vspace{-.25cm}

\subsection{MRTA-CT as Optimization Problem}
\label{sec:mrta_optimization}
\vspace{-.1cm}
The exact solution to the MRTA-CT problem, excluding the communication constraints, can be obtained by formulating it as an MINLP problem, which can be concisely expressed as (for brevity):
\begin{align}
\vspace{-.5cm}
\label{eq:objectiveFunction}
 \min~ f_\text{cost} = (N - N_\text{success})/{N} \\ 
   N_{\text{success}} = \sum_{i \in V} \rho_{i} \nonumber
    \begin{cases}
     \rho_{i} = 1, \ if \ \tau^{f}_{i} \leq \tau_{i} \\
     \rho_{i} = 0, \ if \ \tau^{f}_{i} > \tau_{i} 
    \end{cases}
\end{align}\vspace{-.6cm}
\begin{align}
\label{eq:constraint}
& 0 \leq \Delta_{r}^{t} \leq \Delta_{\text{max}},  r \in R \\
\label{eq:constrain2}
& 0 \leq c_{r}^{t} \leq C_{\text{max}} ,  r \in R
\vspace{-.3cm}
\end{align}
Here $\tau_{i}^{f}$ is the time at which task $i$ is completed, $\Delta_{r}^{t}$ is the available range for robot $r$ at a time instant $t$, 
$c_{r}^{t}$ is the capacity of robot $r$ at time $t$, $N_\text{success}$ is the number of successfully completed tasks during the operation. Here, we craft the objective function (Eq.~\eqref{eq:objectiveFunction}) such that it emphasizes maximizing the completion rate (i.e., the number of completed tasks divided by the total number of tasks). Equations \ref{eq:constraint} and \ref{eq:constrain2} correspond to the remaining range and capacity respectively at time $t$.
To learn policies that yield solutions to this CO problem, we express the MRTA-CT as an MDP over a graph, as described next. 
 \vspace{-.2cm}
\subsection{MDP over a Graph}\label{sec:ProblemFormulation}
\vspace{-.1cm}

The task space of an MRTA-CT problem can be represented as a graph, including a set of nodes/vertices ($V$) and a set of edges ($E$) that connect the vertices to each other. The complete graph is given by $\mathcal{G} = (V, E, \Omega)$, where $\Omega$ is a weighted adjacency matrix. Each node represents a task, and each edge connects a pair of nodes. For MRTA-CT with $N$ tasks, the vertices and edges are $N$ and $N(N-1)/2$, respectively. Node $i$ is assigned a 4-dimensional feature vector denoting the task location coordinates, time deadline, and the remaining workload/demand i.e., $\delta_i$=$[x_i,y_i,\tau_i,w^t_i]$ where $i \in [1,N]$. Here, the weight between two edges $\Omega^{i,j}$ ($\in \Omega$) can be computed as $\Omega^{i,j}$ = $1 / (1+\sqrt{(x_{i}-x_{j})^{2} + (y_{i}-y_{j})^{2} + (\tau_i - \tau_j)^2 + (w^t_i - w^t_j)^2})$, where $i,j \in [1,N]$.

The MDP defined in a decentralized manner for each individual robot (to capture its task selection process) can be expressed as a tuple $<\mathcal{S}, \mathcal{A}, \mathcal{P}_a, \mathcal{R}>$. The components of the MDP can be defined as \textbf{State Space ($\mathcal{S}$)}, i.e., a robot $r$ at its decision-making instance uses a state $s\in\mathcal{S}$, which contains the following information: 1) Task graph $\mathcal{G}$, 2) the current operation time $t$, 3) current destination of robot ($x^{t}_{r}, y^{t}_{r}$), 4) remaining range (battery state) of robot $r$ $\delta^{t}_{r}$, 5) capacity of robot $r$ $c^{t}_{r}$ , 6) destination of its peers ($x_{k}, y_{k}, k \in R \ k \neq r$), 7) the remaining range of peers $\delta^{t}_{k}, k \in R \ k \neq r$, 8) capacity of peers $c^{t}_{k}, k \in R \ k \neq r$, and 9) next decision time of peers $t^{\text{next}}_{k}, k \in R \ \neq r$, 10) the time at which each peer robot took it's previous decision $ts_{k}, k \in R \ k \neq r$. When a robot $k$ visits a task $i$ the demand fulfilled by the robot $k$ is $\text{min}(w^t_i, c^{t}_{k})$.
\noindent\textbf{Action Space ($\mathcal{A}$):} The set of actions is represented as $\mathcal{A}$, where each action $a$ is defined as the index of the selected task, $\{0,\ldots,N\}$ with the index of the depot as $0$. The task $0$ (the depot) can be selected by multiple robots, but the other tasks can be chosen if they are active (not completed or missed tasks). \noindent\textbf{$\mathcal{P}_a(s'|s,a)$:} A robot taking action $a$ at state $s$ reaches the next state $s'$ in a deterministic manner. \textbf{Reward ($\mathcal{R}$):} The reward function is defined as $-f_\text{cost}$, calculated when there are no more active tasks (all tasks demand has been met). \textbf{Transition:} The transition is an event-based trigger. An event is when a robot reaches its selected task or visits the depot location. Here we do not consider any uncertainty, hence the state transition probability is 1. 

\textbf{Communication modeling:} The state variables 6, 7, 8, 9, and 10 are the values that robot $r$ has about its peers during the decision-making time $t$, 
along with a vector ($w_{r} \in \mathbb{R}^N$) which tracks the task completion ratio, 
and gets updated only during an information exchange (about $(6\times M +N)\times 8$ bytes exchanged for a \texttt{double} data type) based on the latest value of $ts_{k}, k \in R, k \neq r$. In our study, two robots can only communicate if their separation distance is less than a threshold distance denoted as $d_{\text{com}}^{\text{thresh}}$. 
\vspace{-.2cm}
\section{Learning Framework}

\label{sec:Learning_Framework} \vspace{-.1cm}
By formulating the MRTA-CT as an MDP, we can use an RL algorithm to learn policies that maximize the objective function (Eq. \ref{eq:objectiveFunction}). The learning framework mainly consists of the CAPAM-TD policy network and a policy gradient RL algorithm. The RL algorithm used here is PPO. The CAPAM policy network from \cite{Paul_ICRA} has demonstrated both generalizability and scalability capabilities and hence we adopt this network and enhance it with TD, with the aim to use higher dimensional topological features for decision-making to maximize the reward. In this section, we will describe the importance of the local higher-order topological information, TD using Persistent Homology, a quick overview of the CAPAM architecture, and how TD is incorporated into the CAPAM network. The CAPAM policy network consists of a Graph Capsule Convolutional Network \cite{Verma2018} based encoder, a context (that reads in robot states), and a Multi-Head Attention (MHA) based decoder \cite{Kool2019}. The encoder takes in the task graph $\mathcal{G}$, computes a feature vector $F_{0i}$ for each graph node $i \in V$ by a linear transformation of the node properties $\delta_i$, which is then passed through multiple Graph Capsule Layers to compute permutation invariant node embeddings. \vspace{-.2cm}
\begin{equation}
    \label{Eq:GraphCapsuleFunction}
    f_{p}^{(l)} (X, \mathcal{L}) = \sigma(\sum_{k=0}^{K}\mathcal{L}^{k}(F_{(l-1)}(X,\mathcal{L})^{\circ p})W_{pk}^{(l)}) \vspace{-.2cm}
\end{equation} 
\noindent Here $\mathcal{L}$ is the graph Laplacian, $p$ is the order of the statistical moment, $K$ is the degree of the convolutional filter, $F_{(l-1)}(X,\mathcal{L})$ is the output from $(l -1)$-th layer, $F_{(l-1)}(X,\mathcal{L})^{\circ p}$ represents $p$ times element-wise multiplication of $F_{(l-1)}(X,\mathcal{L})$. 
Here, $F_{(l-1)}(X,\mathcal{L}) \in \mathbb{R}^{N \times h_{l-1}p}$, $W_{pk}^{(l)} \in \mathbb{R}^{h_{l-1}p \times h_{l}}$. The variable $f_{p}^{(l)} (X, \mathcal{L}) \in \mathbb{R}^{N \times h_{l}}$ is a matrix where each row is an intermediate feature vector for each node $i \in [1,N]$, infusing nodal information from ${L}_{e} \times K$ hop neighbors, for a value of $p$. The output of layer $l$ is obtained by concatenating all $f_{p}^{(l)} (X, \mathcal{L})$, as given by: 
\vspace{-.17cm}
\begin{equation}
    \label{Eq:LayerOutput}
    F_{l}(X,\mathcal{L}) = [f_{1}^{(l)} (X, \mathcal{L}), f_{2}^{(l)} (X, \mathcal{L}), ...f_{P}^{(l)} (X, \mathcal{L})] \vspace{-.1cm}
\end{equation}

Here $P$ denotes the highest order of statistical moment, and $h_{l}$ denotes the node embedding length of layer $l$. We consider all the values of $h_{l}$ (where $l \in [0, L_{e}]$) to be the same for this paper.
Equations \ref{Eq:GraphCapsuleFunction} and \ref{Eq:LayerOutput} were computed for $L_{e}$ layers, where each layer uses the output from the previous layer ($F_{l-1}(X,\mathcal{L})$).
The context reads in the state of the robot taking the decision as well as the state information available to it regarding the peer robots. and computes a vector $Q$ during a decision-making instance. The node embeddings from the encoder $F_{L_{e}}(X,\mathcal{L})$ and the context $Q$ are then passed to the MHA-based decoder, which computes output probabilities for all the nodes. This output probability distribution is used to choose the next task node/task to visit by the robot taking the decision. A complete description of the CAPAM architecture can be found in our previous work \cite{Paul_ICRA}.
\vspace{-.2cm}

\subsection{Topological data analysis using persistent homology}\label{subsec:TDA_using_PH}
\begin{figure}[!ht]
    \centering
    \vspace{-.45cm}
    \includegraphics[width=0.4\textwidth]{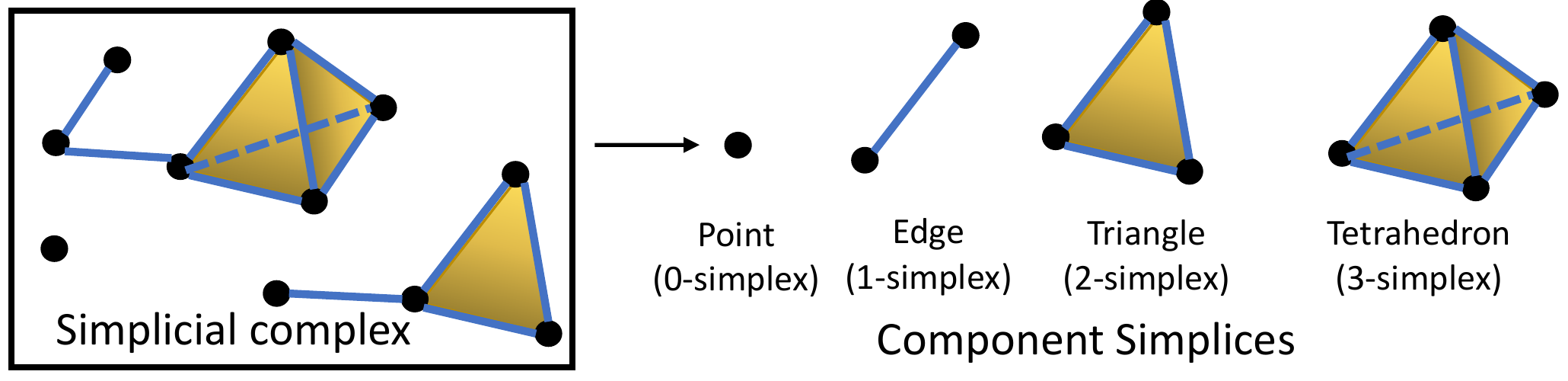}
    \vspace{-0.2cm}
    \caption{A simplicial complex composed of a point, edges, triangles, and tetrahedron.}
    \label{fig:simplicial_complex}
    \vspace{-0.4cm}
\end{figure}
Over the past few years, there have been numerous studies demonstrating the advantage of complementing ML with topological information extracted using the machinery of TDA~\cite{pun2018persistent, townsend2020representation,cang2018integration}, where the underlying structure of data (such as a point cloud or graph) can be used for improving the learning performance. 
In this work, we use \textit{Persistent Homology} (PH) \cite{aktas2019persistence} to extract higher-order topological information from the task graph, in the form of \textit{Persistence Diagram} (PD). We provide here a very brief description of PH and PD. A \textbf{simplicial complex} is an entity composed of a set of points, line segments, triangles, tetrahedron, and their higher dimensional counterparts. The components of a simplicial complex are called simplicies. 

\begin{figure}[!ht]
    \centering
    \vspace{0.1cm}
    \includegraphics[width=0.41\textwidth]{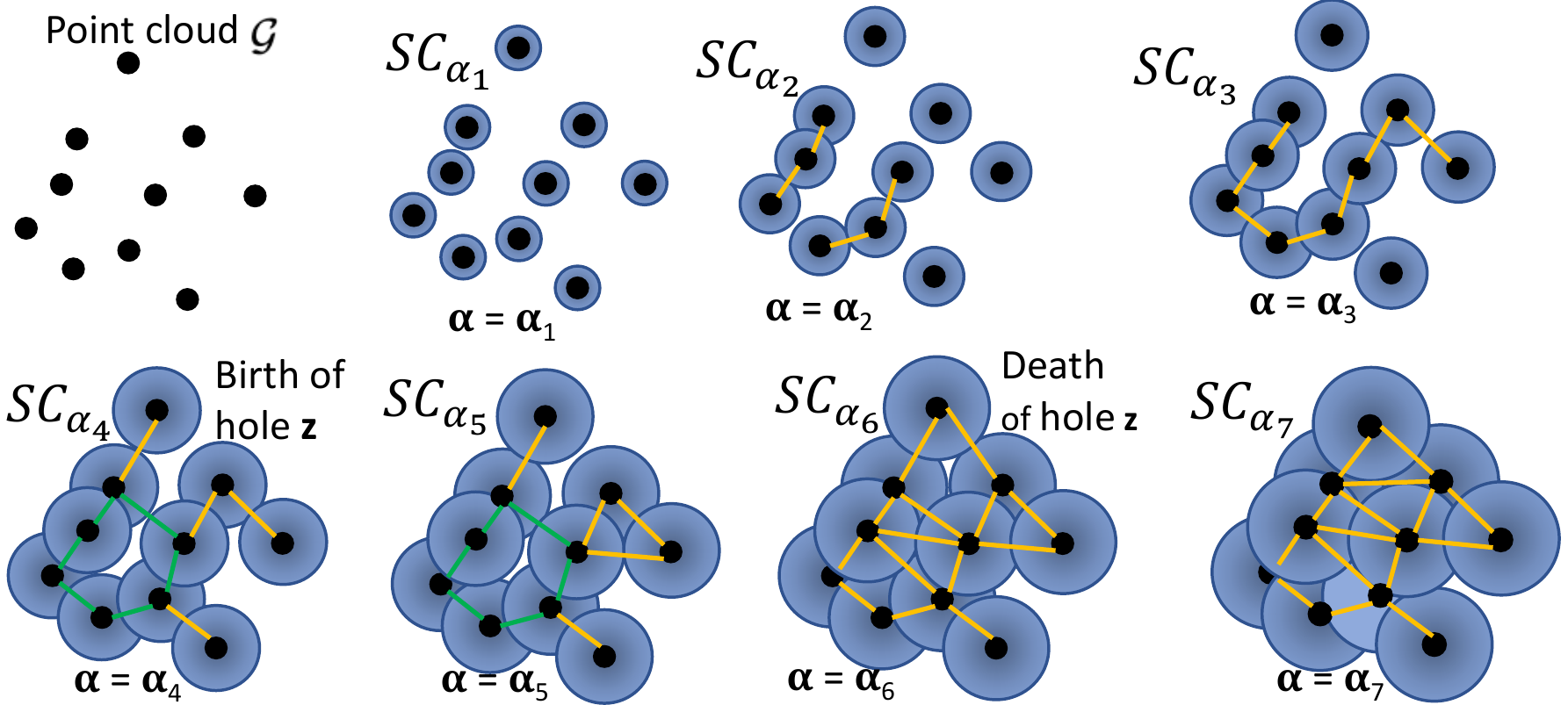}
    \vspace{-0.1cm}
    \footnotesize
    \caption{PH applied to a point cloud $\mathcal{G}$. A $p$-dimensional hole z is characterized by $\alpha$ values during it's birth and death $(\alpha_{4}, \alpha_{6})$. The set which computes the birth and death $\alpha$ for all $p$-dimensional gives PD ($\mathcal{G}$)}
    \label{fig:PH}
    \vspace{-0.2cm}
\end{figure}
\begin{figure}[!ht]
\vspace{-.1cm}
    \centering
    \includegraphics[width=0.44\textwidth]{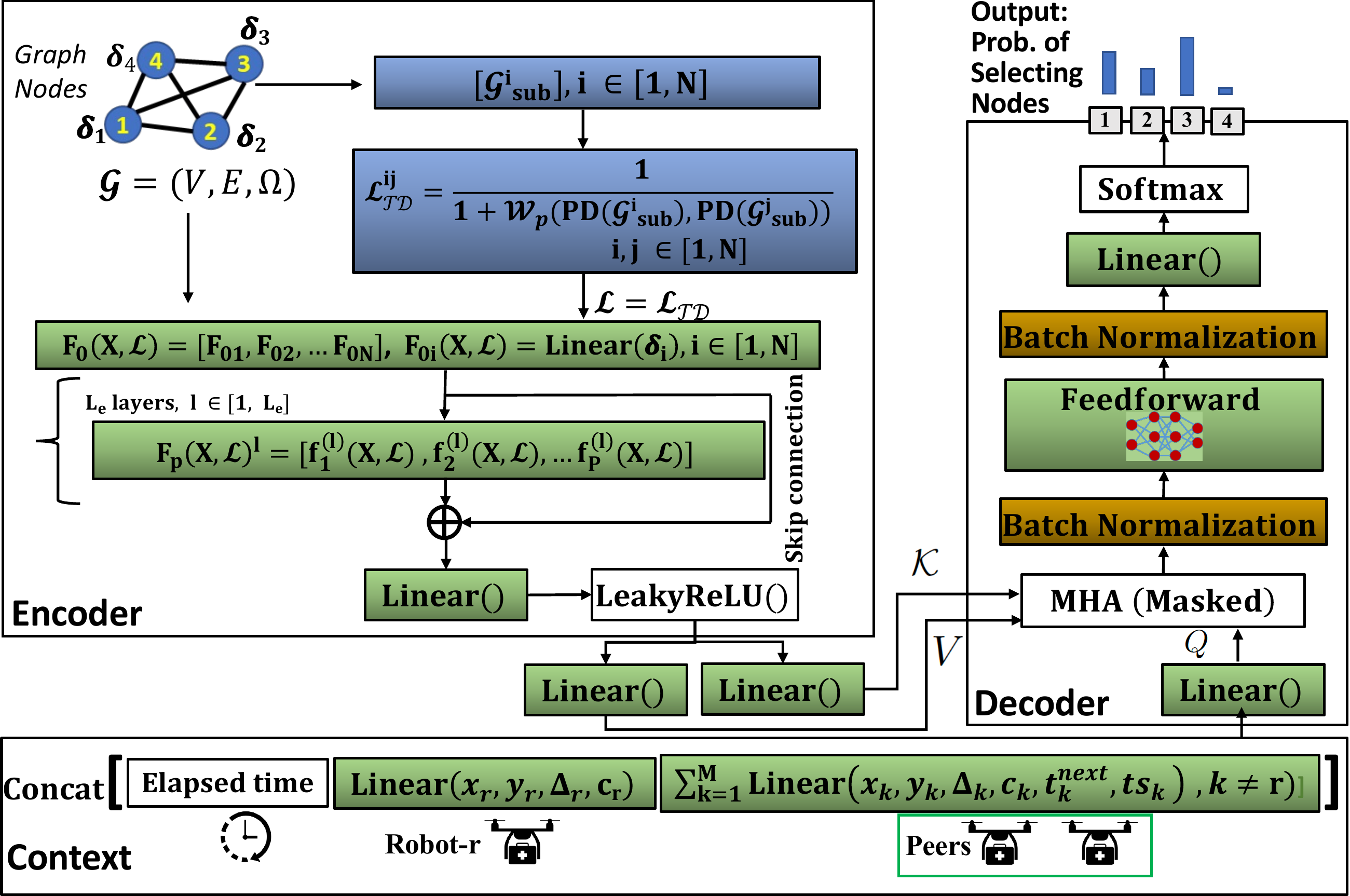}
    \vspace{-0.1cm}
    \footnotesize
    \caption{CAPAM-TD policy network. Blocks in green color have learnable weights. Blocks in blue represent the TD.}
    \label{fig:capam_arch}
    \vspace{-0.5cm}
\end{figure}
For example, given a graph $\mathcal{G} = (V,E,\Omega)$, PH starts with filtration of $\mathcal{G}$ to get a new graph $\mathcal{G_{\alpha}} = (V_{\alpha},E_{\alpha},\Omega_{\alpha})$, where $V_{\alpha}=V$, $(i,j) \in E$ if $\Omega_{\alpha}^{i,j}=1$, and $\Omega_{\alpha}^{i,j}=1$ if $|\delta_{i} - \delta_{j}| \leq \alpha$ else $\Omega_{\alpha}^{i,j}=0$, and a simiplicial complex $\mathcal{SC_{\alpha}}$ can be generated for $\mathcal{G_{\alpha}}$ (Fig. \ref{fig:PH}). Then we identify the occurrence of topological features such as cycles, cavities, and higher $p$-dimensional holes in $\mathcal{SC_{\alpha}}$ and track their lifespans. That is, as the value of $\alpha$ increases (starting from 0), different simplicial complexes are generated which results in the birth of different $p$-dimensional holes and their death. Let $\mathcal{Z}$ be the set of all the $p$-dimensional holes encountered with varying values of $\alpha$, we track the birth and death of the $p$-dimensional holes using their corresponding $\alpha$ values. The longer the lifespan of a $p$-dimensional hole, the likelier this topological feature contains some latent information on the underlying object structure.  All extracted topological features can be summarized in a form of PD. A PD for $\mathcal{G}$ is defined as $PD(\mathcal{G}) = \{{(\alpha^{z}_{1}, \alpha^{z}_{2})}\in \mathbb{R}^2, \forall z \in \mathcal{Z}, \alpha^{z}_{1} < \alpha^{z}_{2}\}$. \vspace{-.2cm}
\subsection{Incorporating TDA with CAPAM}\label{subsec:TDA_with_CapAM}\vspace{-.15cm}
Inspired from \cite{chen2021topological}, to include the local higher-order topological information, we replace the graph Laplacian $\mathcal{L}$, with a Laplacian matrix computed using PH ($\mathcal{L}_{\text{TD}} \in \mathbb{R}^{N \times N}$). First, for each node $i \in V$ of graph $\mathcal{G}$, we find a sub-graph $\mathcal{G}_{\text{sub}}^{i}$ which consists of $k$-hop neighbors of node $i$ (where $k \geq 1$). A node $j \in V$ is an immediate neighbor of node $i$ if $|\delta_{i} - \delta_{j}|$ $<$ $d_{\text{thresh}}$, where $||$ represents the $L_2$ norm and $d_{\text{thresh}}$ is a threshold distance. Hence for each node $i \in V$ we can get a set of points $\mathcal{G}_{\text{sub}}^{i}$ (or sub-graph). For each $\mathcal{G}_{\text{sub}}^{i}, i \in V$ we compute it's Persistence Diagram $PD(\mathcal{G}_{\text{sub}}^{i})$. We use the Wasserstein distance (as in \cite{chen2021topological}) ($\mathcal{W}_p(\cdot)$) (where $0 \leq p \leq \infty$) between each node's PD to compute each element of $\mathcal{L}_{\text{TD}}$, represented as:
\vspace{-.3cm}
\begin{equation}
    \mathcal{L}^{ij}_{\text{TD}} = \frac{1}{(1 + \mathcal{W}_p(PD(\mathcal{G}_{\text{sub}}^{i}), PD(\mathcal{G}_{\text{sub}}^{j})))}, \ \forall \ i,j \in V
\end{equation}
Hence for each pair of nodes $i,j \in V$, $\mathcal{L}^{ij}_{\text{TD}}$ will be close to 1 if they have similar topological information, and close to 0 otherwise. Figure~\ref{fig:capam_arch} shows the overall CAPAM-TD network.
\vspace{-.4cm}

\section{Experimental Evaluation}
\label{sec:Experimental_Evaluation}
\vspace{-.2cm}
\subsection{Baseline methods}\label{subsec:Baseline_methods}\vspace{-.1cm}
We use five baselines for performance comparison of CAPAM-TD-RL on the $\%$ task completion and total time spent for computing the decisions, with simulation time not included. These baselines include: \textbf{1) Mixed Integer Non-Linear Programming (MINLP):} We formulated the MRTA-CT as MINLP and used Gurobi \cite{gurobi} for solving the MINLP. Given that the MINLP solution does not consider the communication constraints, it can be used here to compute the upper bound of the optimality gap of the other methods. In other words, a successfully converged MINLP here finds ideal solutions that are as good as or better than the true optimum solutions. \textbf{2) Bi-Graph MRTA (BIGMRTA)}: BIGMRTA~\cite{ghassemi2019decmrta} is an online method that uses a bipartite graph to connect robots to tasks based on an incentive model. The model considers the task's features and the robot's states to determine the weights of connecting edges, which allows for the decomposition of the problem and yields a measure of robot-task pairing suitability. Each robot solves a maximum weighted matching problem to identify optimal task assignments that maximize the team's net incentive. \textbf{3) Feasibility-preserving Random-Walk (FEASRND)}: FEASRND is a myopic decision-making method that takes randomized but feasible actions, avoiding conflicts and satisfying other problem constraints. \textbf{4) Multi-Layer Perceptron based RL (MLP-RL)}: Here the node encoder of the policy network is a Multi-Layered Perceptron with 2 hidden layers (with 512 neurons), $h_{l}$ =128. \textbf{5) Capsule-Attention Mechanism-based RL (CAPAM-RL)}: Here we use the CAPAM policy network from \cite{Paul_ICRA} with parameters as follows: $K$=3, $L_{e}$=3, $P$=3, $h_{l}$=128, (these parameters are the same for CAPAM-TD).
We consider MINLP and BIGMRTA to be the upper bound performance (even though the computational time is significantly high) and FEASRND to be the lower bound.

\textbf{Environment \& Training details:} We consider a scenario (described in section \ref{sec:Multi_Robot_collective_Transport}) with a single depot, $N$=50, $M$=6 over an area of $1$ $\times$ $1$ ${\text{km}}^{2}$ area. Each robot has maximum payload capacity $C_{\text{max}}$=5kg and the demand for each task $i$ ($w_{i}$) is drawn from a uniform distribution between 1 and 10 kg. Each robot $r \in R$ has a uniform speed of $10$m/s. The time deadline for all the tasks ($\tau_{i}$, $i$ $\in$ $V$) follows a uniform distribution between 150 and 600 seconds. The maximum range for the robots $\Delta_{\text{max}}$=4km. The communication threshold range $d_{\text{com}}^{\text{thresh}}$=100m. 

The simulation environment is developed in \textit{Python} as an \textit{Open AI gym} environment. The three policy networks -- CAPAM-TD, CAPAM, and MLP -- are trained (ensuring convergence) with the same parameters: e.g., Total steps=4$\times10^{6}$, Rollout buffer=4$\times10^{4}$, Batch size=4$\times10^{3}$, Step size=1$\times10^{-6}$, and Entropy coefficient = 0.01) for a fair comparison, on two GPUs (NVIDIA Tesla V100) with 16GB RAM using PPO from \textit{Stable-Baselines3} \cite{stable-baselines3}. To compute PD, we use the \textit{Gudhi} library \cite{gudhi:urm}. The trained models are tested and the non-learning (MINLP, BIGMRTA, FEASRND) based methods are implemented on a 2.6 GHz Intel core i7 MacOS 11.2.3 system. \vspace{-.1cm}

\subsection{Generalizability}\label{subsec:Generalizability}\vspace{-.1cm}
In this paper, \textit{generalizability} refers to the performance of the trained model on unseen test scenarios with the same (or lower) number of tasks as that of the training scenarios; and where the test and training scenarios are drawn from the same probability distribution over task locations, deadlines and demand. In this work, generalizability is evaluated on test scenarios with the number of tasks multiplied by a factor $\lambda_{t}$  $\in$ $\{0.5, 1.0\}$ and drawn from the same distribution as that of training. For each value of $\lambda_{t}$, we consider 3 cases with varying numbers of robots representing small, medium, and large. This gives us a total $2\times3$=$6$ scenario. For each scenario, the number of tasks $N$=$int(\lambda_{t}\times50)$ and the number of robots $M$=$int(6\times \lambda_{t} \times \lambda_{r})+1$, where $\lambda_{r}$ takes a value of 0.5 for small, 1 for medium, and 2 for large number of robots. For each combination of $N$ and $M$ we consider 100 test samples to compare the performance, where all the 100 samples are the same for all the methods.
\begin{figure}[!ht]
\vspace{-.2cm}
    \centering
    \begin{subfigure}{\linewidth}
    \centering
    \includegraphics[width=\textwidth, height=.48\textwidth]{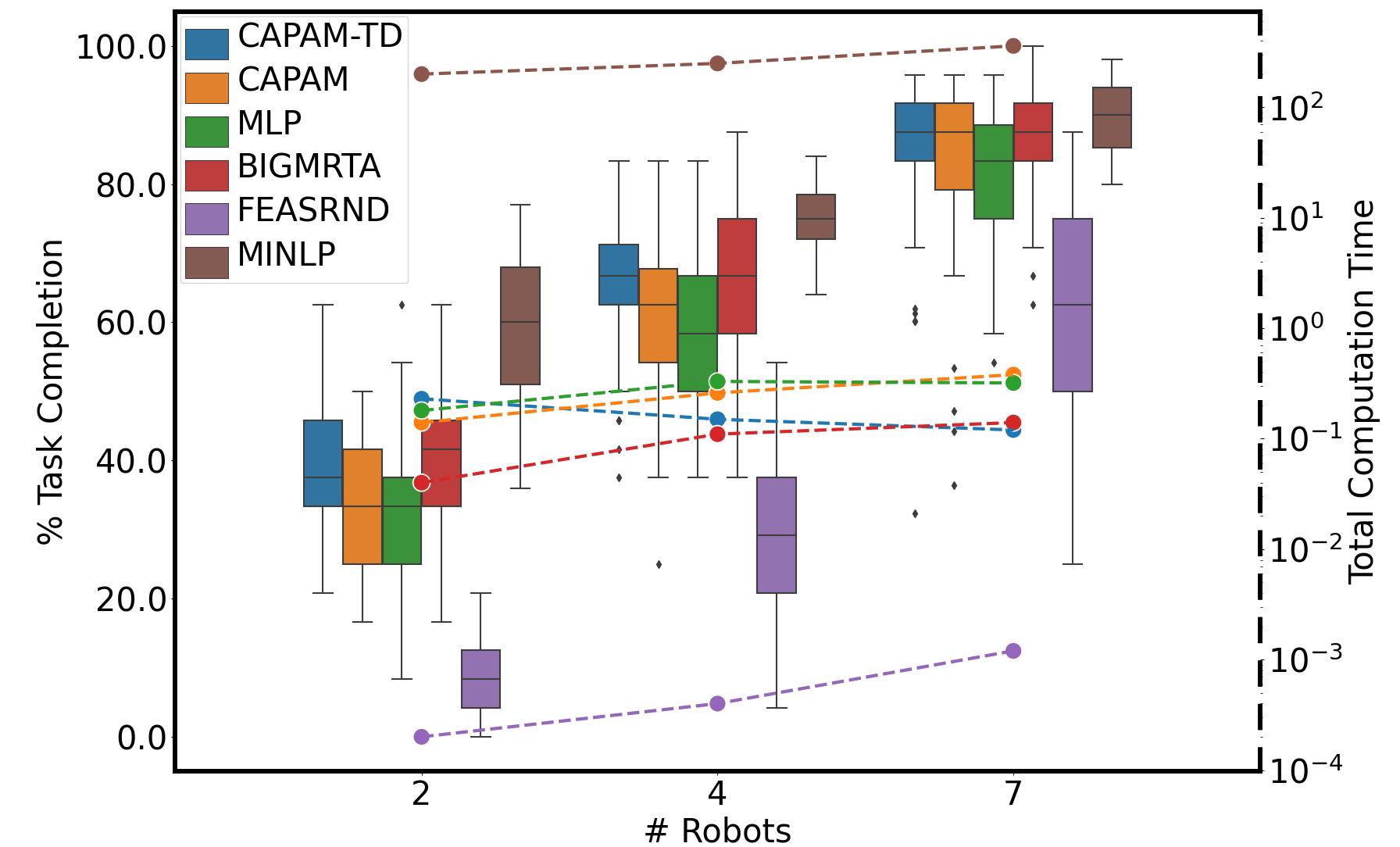}
    \vspace{-.6cm}
    \caption{Scenarios with $N=25$}
    \label{fig:Gen_25}
    \end{subfigure}
    \begin{subfigure}{\linewidth}
    \centering
    \includegraphics[width=\textwidth, height=.48\textwidth]{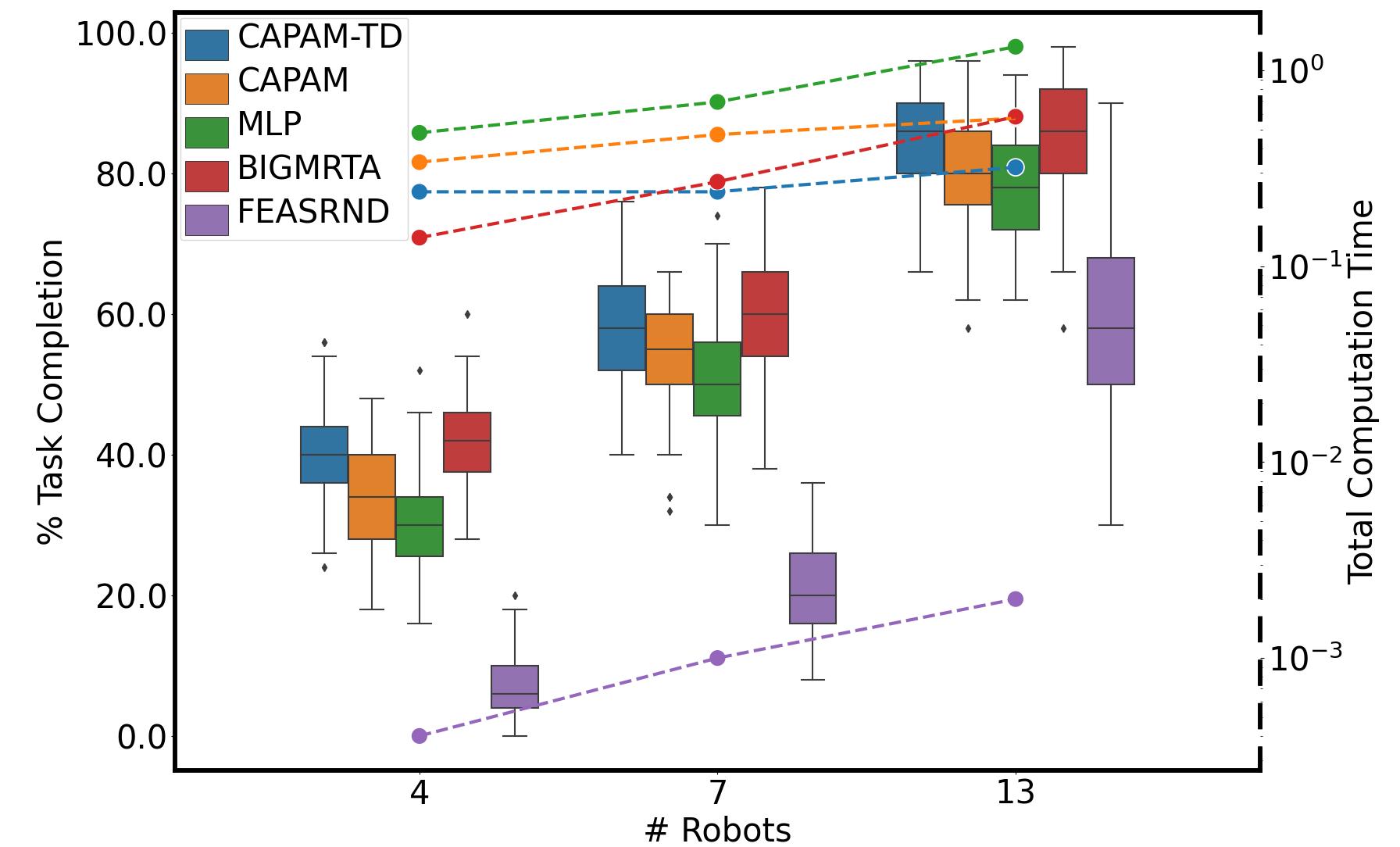}
        \vspace{-.6cm}
    \caption{Scenarios with $N=50$}
    \label{fig:Gen_50}
    \end{subfigure}
        \vspace{-.3cm}
    \caption{$\%$ task completion (box plots to left axis) and average computation time in seconds (dashed lines to the right axis) for generalizability.}
    \label{fig:Gen}
    \vspace{-.4cm}
\end{figure}

As expected, MINLP produced the best results for all the scenarios with $N$=25. However, this comes with a very high computational cost, where the samples (total 100 per scenario) were run for 200, 250, and 360 seconds for $M$=2,4 and 7 respectively, and without considering any communication constraints. Due to high computational expense, the MINLP solution is generated only for scenarios with $N$=25. For larger values of $N$ ($\geq50$), the computational time per sample is very high ($>1000$ seconds per sample). For scenarios with $N$=25 and $N$=50 (Fig.~\ref{fig:Gen}), CAPAM-TD-RL outperforms CAPAM-RL, MLP-RL, and FEASRND. This is confirmed with a statistical t-test with 5\% significance, $p$-value$<0.05$  (except in the case of CAPAM-RL $N$=25, $M$=7). CAPAM-RD-RL also shows a $\%$ task completion performance comparable to BIGMRTA ($p$-value$>0.05$ for all scenarios). 

Even though BIGMRTA has a lower computation time for smaller-sized problems (($N$=25, $M$=2,4), and ($N$=50, $M$=4)), for larger number of robots the computation time is higher. It is important to note the trend of the computation time. In scenarios with $N$=50, the computation time increases from 0.14 ($M$=4) to 0.58 ($M$=13) seconds for BIGMRTA, while for CAPAM-TD-RL the increase is from 0.24 to 0.38 seconds. This difference in the increase of computational time becomes more drastic when scaled to scenarios with larger $N$ and $M$ values, which is discussed in section (\ref{subsec:Scalability}). CAPAM-TD-RL outperforms both learning-based methods (CAPAM-RL and MLP-RL) in terms of the average $\%$ task completion by a maximum margin of 6.5$\%$ (for $N$=25, $M$=2) and 9.9$\%$ (for $N$=50 and $M$=4) respectively. The improved performance of CAPAM-TD-RL compared to CAPAM-TD for generalizability can be credited to the use of PD to compute the graph Laplacian.
\vspace{-.2cm}

\subsection{Scalability}\label{subsec:Scalability} \vspace{-.1cm}
\textit{Scalability} refers to the performance of the trained model on test scenarios with higher numbers of tasks and robots compared to the training scenarios. Here $\lambda_{t}$ takes values of 2, 5, and 10, while $\lambda_{r}$ is the same as that used in the  generalizability analysis.
\begin{figure}
    \centering
    \begin{subfigure}{\linewidth}
    \centering
    \includegraphics[width=\textwidth, height=.473\textwidth]{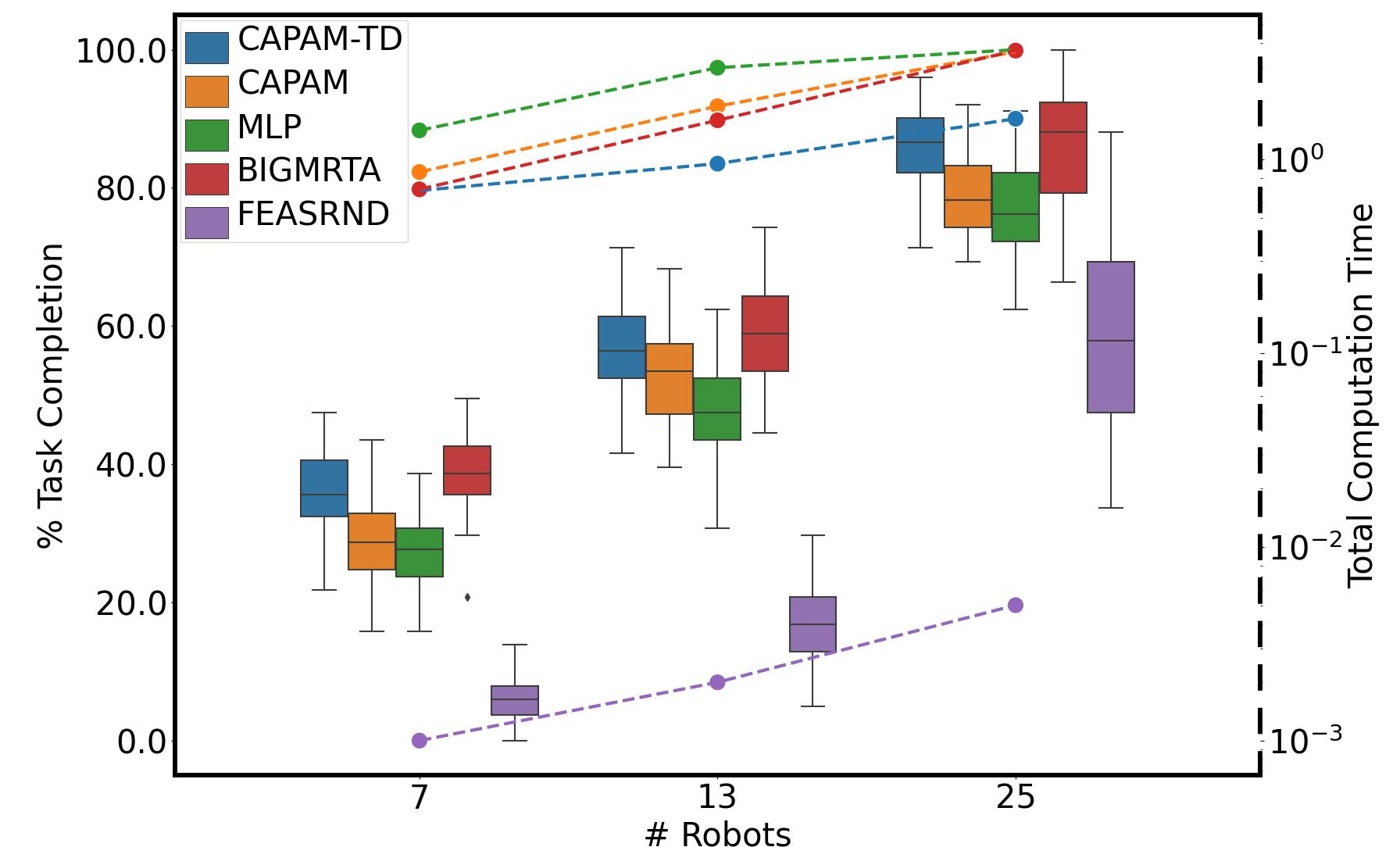}
    \vspace{-.5cm}
    \caption{Scenarios with $N=100$}
    \label{fig:Scal_100}
    \end{subfigure}
    \begin{subfigure}{\linewidth}
    \centering
    \includegraphics[width=\textwidth, height=.473\textwidth]{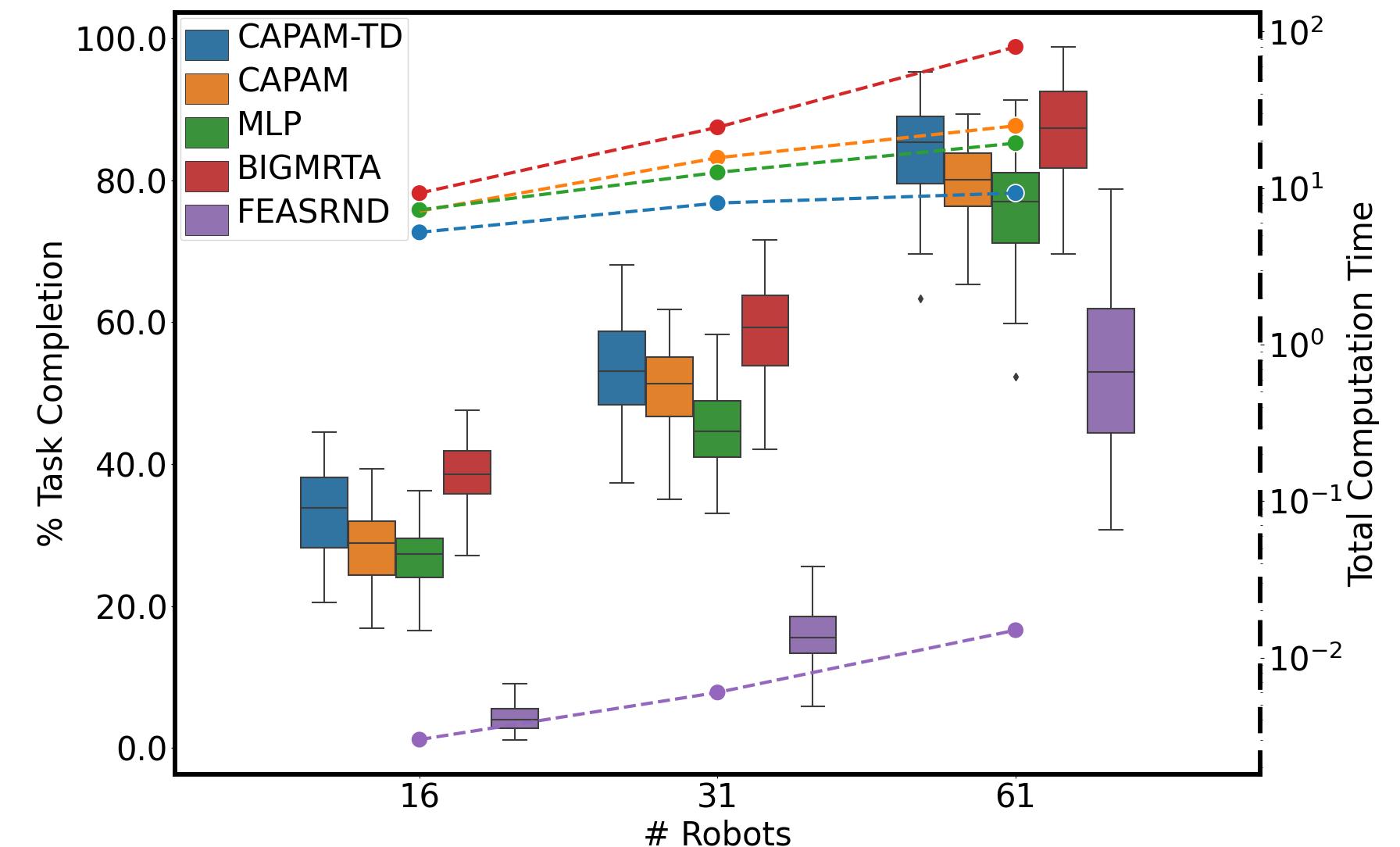}
    \vspace{-.5cm}
    \caption{Scenarios with $N=250$}
    \label{fig:Scal_250}
    \end{subfigure}
    \begin{subfigure}{\linewidth}
    \centering
    \includegraphics[width=\textwidth, height=.473\textwidth]{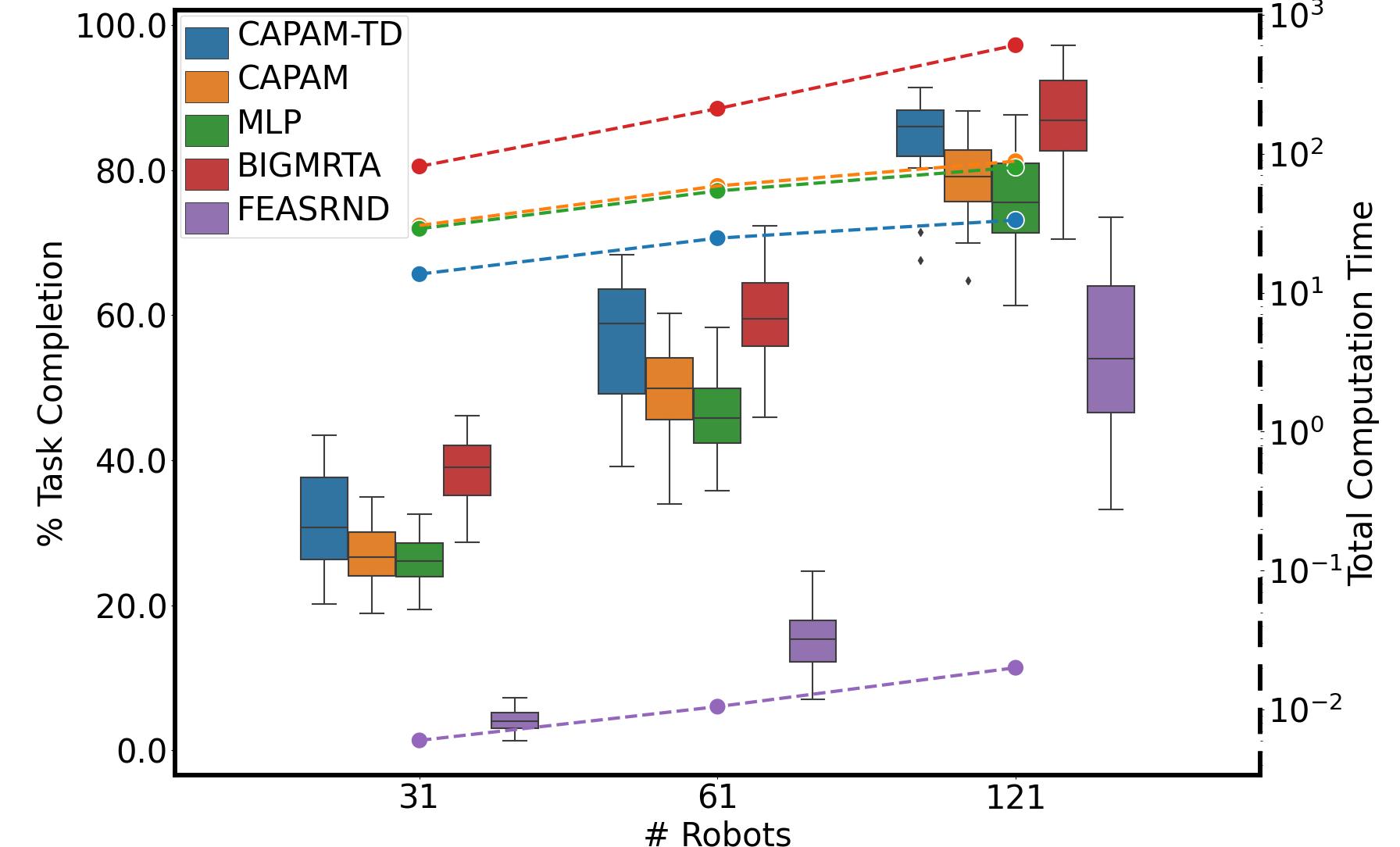}
    \vspace{-.5cm}
    \caption{Scenarios with $N=500$}
    \label{fig:Scal_500}
    \end{subfigure}
    \vspace{-.2cm}
    \caption{$\%$ task completion (box plots to left axis) and average computation time in seconds (dashed lines to the right axis) for Scalability.}
    \label{fig:Scal}
    \vspace{-.7cm}
\end{figure}
The $\%$ task completion follows a similar trend (Fig.~\ref{fig:Scal}) for scalability as seen in the generalizability analysis, where BIGMRTA performs the best and FEASRND performs the worst. CAPAM-TD-RL provides almost comparable performance to BIGMRTA, with the largest performance gap being 6.54 $\%$ ($p$-value=2e-9) for $N$=500, $M$=31, and the smallest being 0.32 $\%$ ($p$-value=0.86) for $N$=100, $M$=25. However, for almost all the scenarios, the computation time of CAPAM-TD-RL is significantly lower compared to BIGMRTA. For scenarios with $N$=500 and $M$=121, the average $\%$ task completion rate for BIGMRTA and CAPAM-TD-RL are 87.2$\%$ and 84.9$\%$ respectively, while their total computation times are 604 and 33 seconds, respectively. Note that BIGMRTA demonstrated comparable performance to MINLP for $N$=25 and $\lambda_{r}$=1,2, with a highest optimiality gap being just 9.4$\%$. This shows that the ability of CAPAM-TD-RL to scale to larger problems without retraining is noticeably better than that of other learning-based methods (CAPAM-RL, MLP-RL). \vspace{-.2cm}

\section{Conclusion}\label{sec:Conclusion}\vspace{-.2cm}
In this paper, we proposed a graph RL approach to generate generalizable, scalable, and real-time executable policies for a class of MRTA problems called MRTA-Collective Transport (MRTA-CT). A particular novelty of our approach lies in the introduction of Topological Descriptors (TD) as additional features that are encoded by the graph neural network serving as the policy model. Persistence homology applied to the task graph  was used to derive the TD features. The proposed GNN, essentially a Capsule Attention Mechanism (CAPAM) based encoder-decoder architecture augmented with TD was trained on randomized problem samples of fixed size in terms of the number of tasks and robots, and was tested on unseen problems of varying size to assess generalizability and scalability. The performance of the CAPAM-TD model was compared to CAPAM without TD, an MLP-based RL, and 3 non-learning-based baseline methods (MINLP, BIGMRTA, and FEASRND). CAPAM-TD-RL demonstrated similar performance compared to BIGMRTA (which itself has comparable performance to MINLP for $N$=25) with roughly 20 times lower computation time, and better performance compared to the other baselines. The advantage of using TD is also readily evident, e.g., in the \{$N$=500, $M$=61\} scalability test, CAPAM with TD achieves 7.1\% better mean completion rate than the one without TD.

\noindent \textbf{Future directions:} As the immediate next step, we intend to extend our method to problems with unreliable communication and task uncertainty, which are common features of the target applications. Further, we plan to implement the CAPAM-TD models on a more realistic multi-robot simulation environment, thereof transitioning to deployment and testing over physical testbeds.




\bibliographystyle{bibliography/IEEEtran}
\bibliography{bibliography/IEEEabrv,bibliography/ICRA}



\section*{APPENDIX}

\subsection{Mixed Integer Non-Linear Programming Formulation for MRTA-Collective Transport Problem}
\label{subsec:MINLP_MRTA_Collective_Transport}
\nomenclature{\(N\)}{Number of tasks}
\nomenclature{\(M\)}{Number of robots}
\nomenclature{\(C_{max}\)}{Maximum capacity of each robot}
\nomenclature{\(\Delta_{max}\)}{Maximum allowed range in a single tour for each robot}
\nomenclature{\((i,j,h,s,r)\)}{Tuple representing a scenario where robot $r$ travels from location $i$ to $j$ during decision number $h$ of route $s$}
\nomenclature{\((h,s,r)\)}{Tuple representing robot $r$ during decision number $h$ of route $s$}
\nomenclature{\((j,h,s)\)}{Tuple representing task $j$ during decision number $h$ of route $s$}
\nomenclature{\(D\)}{Depot}
\nomenclature{\(S\)}{Maximum number of tours for each robot}
\nomenclature{\(R\)}{Set of robots}
\nomenclature{\(T\)}{Set of tasks}
\nomenclature{\(V\)}{Set of tasks including the depot, [D,T]}
\nomenclature{\(H\)}{Maximum number of decisions per tour for each robot}
\nomenclature{\(i,j,k\)}{Indices for nodes}
\nomenclature{\(r\)}{Index for robots}
\nomenclature{\(s\)}{Index for each tour}
\nomenclature{\(h\)}{Index for each decision in a tour}
\nomenclature{\(\tau_{i}\)}{Time deadline for task $i$}
\nomenclature{\(w^{act}_{i}\)}{Demand for task $i$}
\nomenclature{\(t(i,j)\)}{Time to reach from node $i$ to $j$}
\nomenclature{\(d(i,j)\)}{Distance between nodes $i$ and $j$}
\nomenclature{\(x(i,j,h,s,r)\)}{Binary decision variable which takes a value of 1 if robot $r$ travels from node $i$ to $j$ during decision number $h$ of route $s$}
\nomenclature{\(c(h,s,r)\)}{Capacity of robot r after scenario during scenario ($h,s,r$)}
\nomenclature{\(\Delta(h,s,r)\)}{Range of robot r during ($h,s,r$)}
\nomenclature{\(\text{time}(i,j,h,s,r)\)}{time  taken to execute scenario ($i,j,h,s,r$)}
\nomenclature{\(\text{time}^{\text{complete}}_{i}\)}{Time at which task i is completed}
\nomenclature{\(w(j,h,s)\)}{demand met during scenario ($j,h,s$)}
\nomenclature{\(e(i,j,h,s,r)\)}{work done during scenario ($i,j,h,s,r$)}

\printnomenclature

\subsubsection{Objective Function and Constraints}
Objective function eq. \ref{eq:minlp_obj} is set to maximize the number of tasks done, subject to a set of constraints.

\begin{align}
    \label{eq:minlp_obj}
    \max \  \frac{N_{\text{success}} - N}{N}
\end{align}
\begin{align}
    \label{eq:minlp_con_1}
    \sum_{j \in V} x(1,j,1,s,r) = 1,\ \forall s \in [1,S], \ \forall r \in R
\end{align}
\begin{align}
\label{eq:minlp_con_2}
    \sum_{j \in V} x(j,1,H,s,r) = 1,\ \forall s \in [1,S], \ \forall r \in R
\end{align}
\begin{align}
\label{eq:minlp_con_3}
    \sum_{i,j \in V} x(i,j,h,s,r) \leq 1,\ \forall h \in [1,H], \ \forall s \in [1,S], \ \forall r \in R
\end{align}
\begin{align}
\label{eq:minlp_con_4}
    \sum_{j \in V}x(i,j,h,s,r) = \sum_{k \in V} x(k,i,h-1,s,r),\ \\ \nonumber \forall i \in V, \ \forall h \in [2,H] \ \forall s \in [1,S], \ \forall r \in R 
\end{align}
\begin{align}
\label{eq:minlp_con_5}
    \Delta(1,s,r) = \Delta_{max} - \sum_{i,j\in V}x(i,j,1,s,r) \times d(i,j),\ \\ \nonumber \ \forall s \in [1,S], \ \forall r \in R 
\end{align}
\begin{align}
\label{eq:minlp_con_7}
    0 \leq \Delta(h,s,r) \leq \Delta_{max}, \ \forall i,j \in V, \ \\ \nonumber h \in [1, H], \ s \in [1,S], \ r \in R
\end{align}
\begin{align}
\label{eq:minlp_con_22}
    0 \leq c(h,s,r) \leq C_{max}, \ \forall i,j \in V, \ \\ \nonumber h \in [1, H], \ s \in [1,S], \ r \in R
\end{align}
\begin{align}
\label{eq:minlp_con_8}
    \Delta(h,s,r) =  \Delta(h-1,s,r) \\ \nonumber - \sum_{i,j \in V}x(i,j,h,s,r) \times d(i,j), \ \\ \nonumber  \forall h \in [2, H], \ \forall s \in [1,S], \ \forall r \in R
\end{align}
\begin{align}
\label{eq:minlp_con_9}
 w(j,1,1) = \sum_{i \in V, \forall r \in R}e(i,j,1,1,r) \times x(i,j,1,1,r), \ \\ \nonumber \ \forall j \in V
\end{align}
\begin{align}
\label{eq:minlp_con_10}
 w(1,s,r) = w(H,s-1,r) + \\ \nonumber \sum_{\forall i \in V, \forall r \in R} e(i,j,1,s,r) \times x(i,j,1,s,r), \\ \nonumber \forall j \in V, \   \forall s \in [2,S] 
\end{align}
\begin{align}
\label{eq:minlp_con_11}
    0 \leq e(i,j,h,s,r) \leq C_{max}, \ \forall i,j \in V, \  \\ \nonumber \forall h \in [1,H], \forall s \in [1,S], \ \forall r \in R
\end{align}
\begin{align}
\label{eq:minlp_con_12}
    e(i,j,h,s,r) \leq c(h-1,s,r), \ \forall i,j \in V, \  \\ \nonumber \forall h \in [2,H], \forall s \in [1,S], \ \forall r \in R
\end{align}
\begin{align}
\label{eq:minlp_con_13}
    e(i,j,h,s,r) \leq w_{j}^{act} - w(j,h,s), \ \forall i,j \in V, \  \\ \nonumber \forall h \in [2,H], \forall s \in [1,S], \ \forall r \in R
\end{align}
\begin{align}
\label{eq:minlp_con_14}
    w(j,h,s) = w(h-1,s,r) + \\ \nonumber \sum_{\forall i \in V, \forall r \in R}   e(i,j,h,s,r) \times x(i,j,h,s,r), \ \forall j \in V, \\ \nonumber \forall h \in [2,H], \forall s \in [2,S]
\end{align}
\begin{align}
\label{eq:minlp_con_15}
    c(1,s,r) = C_{max} \ - \\ \nonumber  \sum_{i,j\in V}e(i,j,1,s,r) \times x(i,j,1,s,r), \ \\ \nonumber \  \forall s \in [1,S]\,  \forall r \in R
\end{align}
\begin{align}
\label{eq:minlp_con_16}
    c(h,s,r) = c(h-1,s,r) \ - \\ \nonumber \sum_{\forall i,j \in V}  e(i,j,h,s,r) \times x(i,j,h,s,r), \  \\ \nonumber \forall h \in [2,H], \forall s \in [1,S], \ \forall r \in R
\end{align}
\begin{align}
\label{eq:minlp_con_17}
    w(j,h,s) \leq w^{act}_{j}, \ \forall j \in V, \  \\ \nonumber \forall h \in [1,H], \forall s \in [1,S], \ \forall r \in R
\end{align}
\begin{align}
\label{eq:minlp_con_18}
    \sum_{i \in V,  r \in R}w(j,H,S) = w^{act}_{j}, \ \forall j \in V
\end{align}
\begin{align}
\label{eq:minlp_con_19}
    \text{time}(i,j,h,s,r) = t(i,j) \times x(i,j,h,s,r), \ \forall i,j \in V, \  \\ \nonumber \forall h \in [1,H], \forall s \in [1,S], \ \forall r \in R
\end{align}
\begin{align}
\label{eq:minlp_con_20}
    \text{time}^{\text{complete}}_{j} = \sum_{i \in V, h \in [1,H],s \in [1,S],r \in R} \text{time}(i,j,h,s,r), \  \\ \nonumber \forall i,j \in V
\end{align}
\begin{align}
\label{eq:minlp_con_21}
    N_{\text{success}} = \sum_{i \in V-1} \text{Done}_{i} 
    \begin{cases}
     \text{Done}_{i} = 1, \ if \ time^{\text{complete}}_{i} \leq \tau_{i} \\
     \text{Done}_{i} = 0, \ if \ time^{\text{complete}}_{i} > \tau_{i} 
     \\  \forall i \in  V - 1
    \end{cases}
\end{align}

Constraints \ref{eq:minlp_con_1} and \ref{eq:minlp_con_2} ensures that every robot start and ends a route/tour from the depot. Constraint \ref{eq:minlp_con_3} ensures that each robot can make a maximum of one transition during each decision-making instances. Constraint \ref{eq:minlp_con_4} makes the start location of a transition same as the end location of the previous transition for each robot. Constraint \ref{eq:minlp_con_5} sets the range of the robots as $\Delta_{max}$ during the start of a journey. The range update after each transition for the robots are governed by equations \ref{eq:minlp_con_7} and \ref{eq:minlp_con_8}.
Constraints \ref{eq:minlp_con_11}, \ref{eq:minlp_con_12}, and \ref{eq:minlp_con_13} ensures that the work done by a robot do not exceed it's capacity the demand of it's location, while constraints \ref{eq:minlp_con_10} and \ref{eq:minlp_con_14} corresponds to the update for the demand met for each task. Constraints \ref{eq:minlp_con_17} and \ref{eq:minlp_con_18} enforces the the demand met does not exceed the actual demand. Constraints \ref{eq:minlp_con_15} and \ref{eq:minlp_con_16} corresponds to the capacity update for the robots. Constraints \ref{eq:minlp_con_19}, \ref{eq:minlp_con_20}, and \ref{eq:minlp_con_21} are used to compute the task completion time for each task and the number of tasks completed before deadline.

\subsection{More details on CAPAM and PH}\label{subsec:capam_ph}
\subsubsection{CAPAM architecture}\label{subsubsec:CAPAM_arch}
The CAPAM policy architecture proposed in \cite{Paul_ICRA} consists of three parts, which are the encoder, context, and decoder. 

\textbf{Encoder:} The encoder takes in the task information which is represented as a graph $\mathcal{G}$, and computes node embeddings for each node $i \in V$ using the Graph Capsule layers as shown in equations \ref{Eq:GraphCapsuleFunction} and \ref{Eq:LayerOutput} in section \ref{sec:Learning_Framework}.

\textbf{Context:} The context  consists of the following features: \textbf{1)} elapsed mission time; \textbf{2)} range of the robot taking decision; \textbf{3)} capacity of the robot taking decision; \textbf{4)} current location of the robot taking decision; \textbf{5)} current destination of robot's peers; \textbf{6)} range capacity of peer; \textbf{7)} work capacity of peer robots. These features are transformed and aggregated as single learnable vector of length $h_{q}$, which then undergoes a linear transformation to get a vector of length $h_{l}$ also called the query $Q$. It should be noted that features 5,6,and 7 are the state of the peer robots which the robot taking the decision has during the decision-making instance.

\textbf{Decoder:} The MHA-based decoder use the information from the encoder and the context or query, and thereof choose the best task by calculating the probability value of getting selected for each (task) node. In this case, the first step is to feed the embedding for each node (from the encoder) as \textbf{key-values} ($\mathcal{K}$, $V$), since inputs for MHA are key-value pairs \cite{DBLP:journals/corr/VaswaniSPUJGKP17}. The key $K$ and value $V$ for each node is computed by two separate linear transformations of the node embedding  obtained from the encoder.
Now the attention mechanism can be described as mapping the query ($Q$) to a set of key-value ($\mathcal{K},V$) pairs. The inputs, which are the query ($Q$) is a vector, while $\mathcal{K}$ and $V$ are matrices of size $h_{l} \times N$ (since there are $N$ nodes). The output is a weighted sum of the values $V$, with the weight vector computed using the compatibility function expressed as:
\vspace{-.2cm}
\begin{align}
  \label{attention_function}
    \text{Attention}(Q,\mathcal{K},V) = \text{softmax}(Q^{T}\mathcal{K}/\sqrt{h_{l}})V^{T} 
     \vspace{-.2cm}
\end{align} 
where $h_{l}$ is the dimension of the key of any node $i$ ($k_{i} \in \mathcal{K}$).
The output from each MHA layer is obtained as:
\vspace{-.2cm}
\begin{equation}
    \label{mha}
    \text{MHA}(Q,\mathcal{K},V) = \text{Linear}(\text{Concat}(\text{head}_{1}\dots \text{head}_{h_{e}})) 
    \vspace{-.2cm}
\end{equation}
Here $\text{head}_{i} = \text{Attention}(Q,\mathcal{K},V)$ and $h_{e}$ (taken as 8 here) is the number of heads.
A final \texttt{softmax} layer outputs the probability values for all the nodes.  The nodes which are already visited will be masked (by setting their probability as $0$) so that these nodes are not available for selection in the future time steps of the simulation of the multi-robot operation.

\subsubsection{Persistent Homology (PH)}\label{subsubsec:pershom}
The Persistence Diagram (PD) of a point cloud $\mathcal{G}=(V,E,\Omega)$ is computed by applying PH over $\mathcal{G}$. Here we consider $\mathcal{G}$ to be an undirected graph. The next step is to apply a filtration technique to find topological features such as cycles (2-D hole), cavities (3-d hole), and higher dimensional holes. These holes features come into existence as edges are formed between the nodes of the graph. Consider a threshold distance $\alpha$ such that two nodes $u$ and $v$ has an edge between them if $|\delta_{u}-\delta_{v}|\leq \alpha$, where $\delta_{u}$ and $\delta_{v}$ represents the features of nodes $u$ and $v$, respectively. As the value of $\alpha$ increases, new edges are formed which results in new simplical complexes. We track the birth and death of all the k-dimensional holes. For example, from figure \ref{fig:PH}, a 2-d $z$ is born at $\alpha=\alpha_{4}$. This hole persists until $\alpha=\alpha_{6}$ where two new edges are formed and as a result hole $z$ disappears, which is considered at it's death. Therefore $(\alpha_{4}, \alpha_{6})$ denotes the persistence of $z$. Similarly this performed for all the k-dimesional holes $\mathcal{O}$. Therefore the persistence diagram of $\mathcal{G}$ is represented as $PD(\mathcal{G}) = {(\alpha^{1}_{o}, \alpha^{2}_{o}) \forall o \in \mathcal{O}}$, where $\alpha^{1}_{o}$ and $\alpha^{2}_{o}$ represents the birth and death of $o \in \mathcal{O}$.

\subsection{Communication Modelling} \label{subsec:communication_modelling}
In a real world scenario, a full communication between each robots in a mission is not always possible. Each robot will only be able to communicate with robots on close proximity. Robots deployed for disaster response will be implemented with a communication device such as Wifi or Zigbee \cite{8796743} with a typical range of 'x' meters. 
The communication between two robots $a$ and $b$ is two way, where robot `a' shares the information it has about the environment and other robots, to robot `b', and vice versa. This information exchange can be modelled in a probabilistic manner similar to \cite{7063735}, where the information exchange does not happen beyond a threshold distance, and within the threshold distance, the information exchange happens probabilistically, where smaller the separation distance larger is the probability of information exchange. In order to simplify the communication modelling, we assume that information exchange occurs when the separation distance between two robots at a time instant is less than a threshold distance $d_{com}^{thresh}$, and does not occur if the separation distance is greater than $d_{com}^{thresh}$. In this paper, we consider $d_{com}^{thresh} = 100 \ meters$. Apart from the static information such as the location of the tasks and its time deadline, each robot $r$ keeps a record of 1) information regarding the completion of a task $visited_{r}$, 2) information regarding all the robots including (self information) $RobotState_{r}$.

 Each robot $r \in R$ maintains vector ($completion_{r}$) of size $1 \times N$, where each element of $completion_{r}$  corresponds to a task $i$, and represents the fraction of the total demand met for task $i$.
For example, during the information exchange between two robots $a$ and $b$, robot $a$ updates  $visited_{a}$ as, $visited_{a}$ =  $visited_{a}$ $|$ $visited_{b}$, where `$|$' represents a pairwise logical `$\textit{or}$' operator. 

The self state information generated by $r \in R$ includes 1) destination coordinates ($x_{r}$, and $y_{r}$), 2) current available range $\Delta_{r}$ , 3) current payload capacity $c_{r}$ 4) a time stamp when this information was generated $t$.  Let $RobotState_{r}^{l}$ be the state information of robot $l$ that robot $r$ has during an instance.  So we can express $RobotState_{r}^{r} = [x_{r}, y_{r}, \Delta_{r}, c_{r}, t]$. 
 Therefore for all robots $r \in R$, $RobotState_{r}$ is a matrix of size $M \times 5$, where each row $l$ represents $RobotState_{r}^{l}$. 

While information exchange occurs between two robots $a$ and $b$, the state information update happens only for those entries with newer timestamps. For example, during an information exchange between robots $a$ and $b$, robot $a$ has both $RobotState_{a}$ and $RobotState_{b}$, and robot $a$ will replace $RobotState_{a}^{l}$ ($l \in R$) with $RobotState_{b}^{l}$ if $RobotState_{b}^{l}$ has a newer timestamp compared to that of $RobotState_{a}^{l}$.

\end{document}